\documentclass[aps,a4paper,superscriptaddress,preprintnumbers,showpacs,amsmath,amssymb]{revtex4}
\usepackage{graphicx}
\usepackage{dcolumn}
\usepackage{color}
\usepackage{latexsym,amsfonts}

\usepackage{textcomp}

\usepackage{bm}

\usepackage{ulem}

\begin{document}

\title{Einstein--Cartan Gravity with Torsion Field \\ Serving as
  Origin for Cosmological Constant or Dark Energy Density}

\author{A. N. Ivanov}\email{ivanov@kph.tuwien.ac.at}
\affiliation{Atominstitut, Technische Universit\"at Wien, Stadionallee
  2, A-1020 Wien, Austria}
\author{M. Wellenzohn}\email{max.wellenzohn@gmail.com}
\affiliation{Atominstitut, Technische Universit\"at Wien, Stadionallee
  2, A-1020 Wien, Austria} \affiliation{FH Campus Wien, University of
  Applied Sciences, Favoritenstra\ss e 226, 1100 Wien, Austria}

\date{\today}

\begin{abstract}
We analyse the Einstein--Cartan gravity in its standard form ${\cal R}
= R + {\cal K}^2$, where ${\cal R}$ and $R$ are the Ricci scalar
curvatures in the Einstein--Cartan and Einstein gravity, respectively,
and ${\cal K}^2$ is the quadratic contribution of torsion in terms of
the contorsion tensor ${\cal K}$. We treat torsion as an external (or
a background) field and show that the contribution of torsion to the
Einstein equations can be interpreted in terms of the torsion
energy--momentum tensor, local conservation of which in a curved
spacetime with an arbitrary metric or an arbitrary gravitational field
demands a proportionality of the torsion energy--momentum tensor to a
metric tensor, a covariant derivative of which vanishes because of the
metricity condition. This allows to claim that torsion can serve as
origin for vacuum energy density, given by cosmological constant or
dark energy density in the Universe.  This is a model--independent
result may explain a small value of cosmological constant, which is a
long--standing problem of cosmology. We show that the obtained result
is valid also in the Poincar\'{e} gauge gravitational theory by Kibble
(T. W. B. Kibble, J. Math. Phys. {\bf 2}, 212 (1961)), where the
Einstein--Hilbert action can be represented in the same form ${\cal R}
= R + {\cal K}^2$.
\end{abstract}
\pacs{03.65.Pm, 04.20.Cv, 04.60.Bc, 14.80.Va}

\maketitle

\section{Introduction}
\label{sec:introduction}

Torsion is a natural geometrical quantity additional to the metric
tensor. It is accepted
\cite{Hehl1976,Hehl2007,Hehl2012,Hehl2013,Shapiro2002,Hammond2002,Kostelecky2004,Ni2010}
that torsion characterizes spacetime geometry through spin--matter
interactions, which allow to probe the rotational degrees of freedom
of spacetime in terrestrial laboratories
\cite{Rumpf1979,Laemmerzahl1997,Kostelecky2008,Obukhov2014,Lehnert2014,Ivanov2015,Ivanov2015a,Ivanov2015b,Ivanov2016}. However,
as has been shown recently \cite{Ivanov2016}, the requirement of the
linking torsion and fermion spin through torsion--fermion minimal
couplings is violated in the low--energy approximation in curved
spacetimes with rotation (see Eq.(22) of Ref.\cite{Ivanov2016}). The
later allows to admit the existence of torsion even without spinning
matter.  In such an approach torsion can be treated as an external (or
a background) field, defined by a third--order tensor ${\cal
  T}_{\sigma\mu\nu}$, antisymmetric with respect to indices $\mu$ and
$\nu$, i.e. ${\cal T}_{\sigma\mu\nu} = - {\cal T}_{\sigma\nu\mu}$
\cite{Shapiro2002,Kostelecky2008,Lehnert2014,Ivanov2015,Ivanov2015a,Ivanov2015b,Ivanov2016},
which can be introduced into the Einstein--Cartan gravitational theory
as an antisymmetric part of the affine connection through the
metricity condition \cite{Rebhan2012}. Such a torsion tensor field
possesses 24 independent components, which can be decomposed into four
vector ${\cal E}_{\mu} = ({\cal E}_0, - \vec{\cal E}\,)$, four
axial--vector ${\cal B}_{\mu} = ({\cal K}, - \vec{\cal B}\,)$ and
sixteen tensor ${\cal M}_{\sigma\mu\nu}$ components
\cite{Shapiro2002,Kostelecky2008} (see also \cite{Ivanov2015a}). As
has been shown in \cite{Ivanov2015a}, only torsion axial--vector
${\cal B}_{\mu}$ components are present in the torsion--fermion
minimal couplings in the curved spacetimes with metric tensors,
providing vanishing time--space (space--time) components of the
vierbein fields. The torsion vector ${\cal E}_{\mu}$ and tensor ${\cal
  M}_{\sigma\mu\nu}$ components, coupled to Dirac fermions, appear
through torsion--fermion non--minimal couplings with phenomenological
coupling constants \cite{Kostelecky2008} (see also
\cite{Ivanov2015a}). The presence of phenomenological coupling
constants screens real values of torsion vector ${\cal E}_{\mu}$ and
tensor ${\cal M}_{\sigma\mu\nu}$ components. Nevertheless, an
observation of these non--minimal torsion--fermion interactions should
testify an existence of torsion and correctness of the
Einstein--Cartan gravitational theory. It should be emphasized that as
has been shown in \cite{Ivanov2015a} some effective low--energy
interactions of torsion 4--vector ${\cal E}_{\mu} = ({\cal E}_0, -
\vec{\cal E}\,)$ and tensor ${\cal M}_{\sigma\mu\nu}$ components,
caused by non--minimal torsion--fermion couplings, do not depend on a
fermion spin.  Then, as has been shown in
\cite{Ivanov2015b,Ivanov2016}, torsion vector and tensor components
can be probed in terrestrial laboratories through torsion--fermion
minimal couplings in the spacetimes with rotation
\cite{LL2008,Hehl1990,Obukhov2009,Obukhov2011}.  Some steps to
creation of such spacetimes in terrestrial laboratories have been made
by Atwood {\it et al.}  \cite{Atwood1984} and Mashhoon
\cite{Mashhoon1988}, who used rotating neutron interferometers. The
estimates of constant torsion, coupled to Dirac fermions, have been
carried out by L\"ammerzahl \cite{Laemmerzahl1997}, Kostelecky {\it et
  al.} \cite{Kostelecky2008} and Obukhov {\it et al.}
\cite{Obukhov2014} and discussed in \cite{Ivanov2015a}. Recently in
the liquid ${^{4}}{\rm He}$ Lehnert, Snow, and Yan \cite{Lehnert2014}
have measured a rotation angle $\phi_{\rm PV}$ of the neutron spin
about a neutron 3--momentum $\vec{p}$ per unit length $d\phi_{\rm
  PV}/dL$. Using the results, obtained by Kostelecky {\it et al.}
\cite{Kostelecky2008}, Lehnert {\it et al.} \cite{Lehnert2014} have
found that $d\phi_{\rm PV}/dL = 2\zeta$. The parameter $\zeta$ is a
superposition of the scalar $T_0 \sim {\cal E}_0$ and pseudoscalar
$A_0 \sim {\cal K}$ torsion components equal to $\zeta = (2m
\xi^{(5)}_8 - \xi^{(4)}_2)\,T_0 + (2m \xi^{(5)}_9 -
\xi^{(4)}_4)\,A_0$, where $m$ is the neutron mass and $\xi^{(5)}_8$,
$\xi^{(4)}_2$, $\xi^{(5)}_9$ and $\xi^{(4)}_4$ are phenomenological
constants, introduced by Kostelecky {\it et al.}
\cite{Kostelecky2008}. The experiment by Lehnert {\it et al.}
\cite{Lehnert2014} is based on the phenomenon of neutron optical
activity, related to a rotation of the plane of polarization of
transversely polarized slow--neutron beam moving through matter. As
has been reported by Lehnert {\it et al.}  \cite{Lehnert2014}, $\zeta$
is restricted from above by $|\zeta| < 9.1\times 10^{-14}\,{\rm eV}$
at $68\,\%$ of C.L. \cite{Lehnert2014}. Such an estimate is by a
factor $10^5$ larger compared with the upper bound $|\zeta| <
10^{-18}\,{\rm eV}$, calculated in \cite{Ivanov2015a} by using the
estimates by Kostelecky {\it et al.}  \cite{Kostelecky2008}.

In this paper we analyse the Einstein--Cartan gravitational theory
without fermions. The aim of this paper is to show that torsion as a
geometrical characteristic of a curved spacetime additional to a
metric tensor can exist independently of spinning matter and play an
important role in the evolution of the Universe. Torsion in such an
approach is treated as an external (or a background) field
\cite{Shapiro2002,Kostelecky2008,Obukhov2014,Lehnert2014,Ivanov2015,Ivanov2015a,Ivanov2015b,Ivanov2016}. In
section \ref{sec:gravity} we show that the gravity--torsion part of
the Einstein--Hilbert action of the Einstein--Cartan gravitational
theory can be given in the additive form $\int d^4x\,\sqrt{-g}\,{\cal
  R} = \int d^4x\,\sqrt{-g}\,R + \int d^4x\,\sqrt{-g}\,{\cal C}$,
where ${\cal R} = g^{\mu\nu}\,{\cal R}_{\mu\nu}$ and $R =
g^{\mu\nu}R_{\mu\nu}$ are scalar curvatures in the Einstein--Cartan
and Einstein gravity, respectively, with the Ricci tensor $R_{\mu\nu}$
defined in terms of the metric tensor $g_{\mu\nu}$ only
\cite{Rebhan2012}. Then, ${\cal C} = g^{\mu\nu}{\cal C}_{\mu\nu} =
g^{\mu\nu}({{\cal K}^{\varphi}}_{\alpha\mu}\,{{\cal
    K}^{\alpha}}_{\nu\varphi} - {{\cal
    K}^{\alpha}}_{\alpha\varphi}{{\cal K}^{\varphi}}_{\nu\mu})$ is
defined by torsion in terms of the contorsion tensor ${\cal
  K}_{\sigma\mu\nu} = \frac{1}{2}({\cal T}_{\sigma\mu\nu} + {\cal
  T}_{\mu\sigma\nu} + {\cal T}_{\nu\sigma\mu})$ \cite{Kostelecky2004},
and $g = {\rm det}\{g_{\mu\nu}\}$. The raising and lowering of indices
are performed with metric tensors $g^{\mu\nu}$ and $g_{\mu\nu}$,
respectively.  In section \ref{sec:torsion} in a curved spacetime with
an arbitrary metric tensor we derive the Einstein equations in the
Einstein--Cartan gravitational theory with the chameleon
(quintessence) field and matter, defined in the Cold Dark Matter (CDM)
model \cite{PDG2014} in terms of a matter density $\rho$ in the
Einstein frame \cite{Chameleon1,Chameleon2,Brax2004,Ivanov2016a}. The
account for the contribution of the chameleon field
\cite{Chameleon1,Chameleon2} is justified by its property i) to be
responsible for the late--time acceleration of the Universe expansion
\cite{Brax2004,Ivanov2016a} and ii) to have a locally conserved
energy--momentum tensor in a curved spacetime with an arbitrary metric
tensor (see Appendix A).  We show that i) torsion does not couple to
spinless matter and ii) the contribution of torsion to the Einstein
equations can be interpreted in terms of the torsion energy--momentum
tensor $T^{(\rm tors)}_{\mu\nu}$. Since the Einstein tensor
$G^{\mu\nu} = R^{\mu\nu} - \frac{1}{2}\,g^{\mu\nu}\,R$, where $R =
g_{\mu\nu} R^{\mu\nu}$ is the scalar curvature, obeys the Bianchi
identity ${G^{\mu\nu}}_{;\mu} = 0$, where ${G^{\mu\nu}}_{;\mu}$ is a
covariant divergence, in a curved spacetime with an arbitrary metric
tensor $g_{\mu\nu}$ or an arbitrary gravitational field
\cite{Rebhan2012}, the total energy--momentum tensor of the system,
including torsion, the chameleon field and matter, should be also
locally conserved.  We show (see Appendix A) that the energy--momentum
tensor of the chameleon field has a vanishing covariant divergence,
i.e. locally conserved in a curved spacetime with an arbitrary metric
tensor (or an arbitrary gravitational field). Then, we show that the
matter energy--momentum tensor, defined in the CDM model, obeys in a
curved spacetime with an arbitrary metric tensor the evolution
equation, which reduces in the Friedmann flat spacetime to the
evolution equation, derived in \cite{Ivanov2016a}. Because of the
Bianchi identity for the Einstein tensor $G^{\mu\nu}$ and local
conservation of matter and chameleon field energy--momentum tensors in
curved spacetimes with arbitrary metric tensors the torsion
energy--momentum tensor has to be also locally conserved at the same
conditions. Since in our approach torsion is an external (or a
background) field
\cite{Laemmerzahl1997,Kostelecky2008,Obukhov2014,Ivanov2015a,Ivanov2015b,Ivanov2016}
and it is not governed by any equation of motion and boundary
conditions, such a local conservation can be fulfilled if and only if
the torsion energy--momentum tensor is proportional to a metric tensor
$T^{(\rm tors)}_{\mu\nu} \sim g_{\mu\nu}$, which covariant derivative
vanishes because of the metricity condition ${g^{\mu\nu}}_{;\rho} = 0$
\cite{Hehl1976,Kostelecky2004,Rebhan2012}. As a result, the torsion
energy--momentum tensor becomes equivalent to the vacuum
energy--momentum tensor, the contribution of which can be described in
terms of cosmological constant \cite{Rebhan2012} or dark energy
density \cite{Peebles2003,Copeland2006}.  This gives the relation
${\cal C} = g^{\mu\nu}{\cal C}_{\mu\nu} = g^{\mu\nu}({{\cal
    K}^{\varphi}}_{\alpha\mu}\,{{\cal K}^{\alpha}}_{\nu\varphi} -
{{\cal K}^{\alpha}}_{\alpha\varphi}{{\cal K}^{\varphi}}_{\nu\mu}) = -
2\Lambda_C$ (see Eq.(\ref{eq:25})). We would like to emphasize that
the identification of the contribution of torsion to the
Einstein--Hilbert action and to the Einstein equations with the
contribution of cosmological constant is a model--independent because
of a requirement of local conservation in a spacetime with an
arbitrary metric tensor. We may also argue that the constraint
$g^{\mu\nu}({{\cal K}^{\varphi}}_{\alpha\mu}\,{{\cal
    K}^{\alpha}}_{\nu\varphi} - {{\cal
    K}^{\alpha}}_{\alpha\varphi}{{\cal K}^{\varphi}}_{\nu\mu})= -
2\Lambda_C$ admits variations of torsion tensor field as an external
field in sufficiently broad limits of its components. Indeed, such a
constraint looks like a surface in the space of 24 torsion independent
components. Thus, such a torsion--induced cosmological constant is
able to explain a small value of cosmological constant, which is a
long--standing problem of cosmology \cite{Weinberg1989} (see also
\cite{Peebles2003}). Of course, the probes of torsion tensor field
components can be possible only through interactions with spin
particles in particular with Dirac fermions
\cite{Rumpf1979,Laemmerzahl1997,Kostelecky2004,Kostelecky2008,Obukhov2014,Lehnert2014,Ivanov2015,Ivanov2015a,Ivanov2015b,Ivanov2016}.
Nevertheless, we have to emphasize that not all of torsion--fermion
interactions are defined by a fermion spin. As has been shown in
\cite{Ivanov2016} in curved spacetimes with rotation torsion scalar
and tensor components couple to massive Dirac fermions through
low--energy non--spin interactions, caused by minimal torsion--fermion
couplings. In the section \ref{sec:conclusion} we discuss the obtained
results and an equivalence between the Einstein--Cartan gravitational
theory, analysed in this paper, and the Poincar\'{e} gauge
gravitational theory \cite{Kibble1961} (see also
\cite{Sciama1961,Utiyama1956,Blagojevic2001} and
\cite{Hehl1976,Hehl2007,Hehl2012,Hehl2013,Obukhov2014}) without
spinning matter. In Appendix A we calculate the covariant divergence
of the energy--momentum tensor of the chameleon (quintessence) field
and show that it vanishes in a curved spacetime with an arbitrary
metric tensor. In Appendix B we analyse the obtained results in the
Poincar\'{e} gauge gravitational theory, proposed by Kibble
\cite{Kibble1961} (see also
\cite{Sciama1961,Utiyama1956,Blagojevic2001} and
\cite{Hehl1976,Hehl2007,Hehl2012,Hehl2013,Obukhov2014}). We show that
the integrand of the Einstein--Hilbert action $e \,{\cal R} =
e\,{e^{\mu}}_a {e^{\nu}}_b {{\cal R}_{\mu\nu}}^{ab}$ of the
Poincar\'{e} gauge gravitational theory, where $e = \sqrt{- g}$ and
${{\cal R}_{\mu\nu}}^{ab}$ is the gravitational field strength tensor
of the Poincar\'{e} gauge gravitational theory, defined in terms of
the vierbein fields ${e^{\mu}}_a$ and ${e^{\nu}}_b$ and torsion, can
be represented in the additive form $e\,(R + {\cal C})$, where $R =
{e^{\mu}}_a {e^{\nu}}_b {R_{\mu\nu}}^{ab}$ and ${R_{\mu\nu}}^{ab}$ is
the gravitational field strength tensor of the Poincar\'{e} gauge
gravitational theory, defined only in terms of vierbein fields, and
${\cal C} = {{\cal K}^{\varphi}}_{\alpha\mu}\,{{\cal
    K}^{\alpha\mu}}_{\varphi} - {{\cal
    K}^{\alpha}}_{\alpha\varphi}{{\cal K}^{\varphi\mu}}_{\mu}$. This
allows to determine the contribution of torsion to the Einstein
equations through the torsion energy--momentum tensor, local
conservation of which demands its proportionality to a metric tensor.

\section{Einstein--Hilbert action in the Einstein--Cartan gravity with
 torsion and without chameleon field}
\label{sec:gravity}

The Einstein--Hilbert action $S_{\rm EH}$ of the Einstein--Cartan
gravity with torsion we take in the standard model--independent form
\begin{eqnarray}\label{eq:1}
S_{\rm EH} = \frac{1}{2}\,M^2_{\rm Pl}\int d^4x\,\sqrt{-g}\,{\cal R},
\end{eqnarray}
where $M_{\rm Pl} = 1/\sqrt{8\pi G_N} = 2.435\times 10^{27}\,{\rm eV}$
is the reduced Planck mass and $G_N$ is the Newtonian gravitational
constant \cite{PDG2014} and $g$ is the determinant of the metric
tensor $g_{\mu\nu}$. The scalar curvature ${\cal R}$ is defined by
\cite{Kostelecky2004}
\begin{eqnarray}\label{eq:2}
{\cal R} = g^{\mu\nu}{{\cal R}^{\alpha}}_{\mu\alpha\nu} = g^{\mu\nu}
\Big(\partial_{\nu}{\Gamma^{\alpha}}_{\alpha\mu} -
\partial_{\alpha}{\Gamma^{\alpha}}_{\nu\mu} +
        {\Gamma^{\alpha}}_{\nu\varphi}
        {\Gamma^{\varphi}}_{\alpha\mu} -
        {\Gamma^{\alpha}}_{\alpha\varphi}{\Gamma^{\varphi}}_{\nu\mu}\Big)
        = g^{\mu\nu} {\cal R}_{\mu\nu},
\end{eqnarray}
where ${{\cal R}^{\alpha}}_{\mu\beta\nu}$ and ${\cal R}_{\mu\nu}$ are
the Riemann and Ricci tensors in the Einstein--Cartan gravitational
theory, respectively, and ${\Gamma^{\alpha}}_{\mu\nu}$ is the affine
connection
\begin{eqnarray}\label{eq:3}
{\Gamma^{\alpha}}_{\mu\nu} = \{{^{\alpha}}_{\mu\nu}\} + {{\cal
    K}^{\alpha}\,}_{\mu\nu} = \{{^{\alpha}}_{\mu\nu}\} +
g^{\alpha\sigma}{\cal K}_{\sigma\mu\nu}.
\end{eqnarray}
Here $\{{^{\alpha}}_{\mu\nu}\}$ are the Christoffel symbols
\cite{Rebhan2012}
\begin{eqnarray}\label{eq:4}
\{{^{\alpha}}_{\mu\nu}\} =
\frac{1}{2}g^{\alpha\lambda}\Big(\frac{\partial
  g_{\lambda\mu}}{\partial x^{\nu}} + \frac{\partial
  g_{\lambda\nu}}{\partial x^{\mu}} - \frac{\partial
  g_{\mu\nu}}{\partial x^{\lambda}}\Big)
\end{eqnarray}
and ${\cal K}_{\sigma\mu\nu}$ is the contorsion tensor, related to
torsion ${\cal T}_{\sigma\mu\nu}$ by $ {\cal K}_{\sigma\mu\nu} =
\frac{1}{2}\,({\cal T}_{\sigma\mu\nu} + {\cal T}_{\mu\sigma\nu} +
     {\cal T}_{\nu\sigma\mu})$ and ${{\cal T}^{\alpha}}_{\mu\nu} =
     {\Gamma^{\alpha}}_{\mu\nu} - {\Gamma^{\alpha}}_{\nu\mu}$
     \cite{Kostelecky2004}. In case of zero torsion the Riemann and
     Ricci tensors reduce to their standard form
     \cite{Rebhan2012}. The integrand of the Einstein--Hilbert action
     Eq.(\ref{eq:1}) can be represented in the following form
\begin{eqnarray}\label{eq:5}
\sqrt{- g}\,{\cal R} = \sqrt{-g}\,R + \sqrt{-g}\,{\cal C} +
\partial_{\mu}(\sqrt{-g}\,{{\cal K}^{\alpha}}_{\alpha\mu}) -
\sqrt{-g}\,g^{\mu\nu}\Big(\frac{1}{\sqrt{-g}}\,\partial_{\alpha}(\sqrt{-g}\,{{\cal
    K}^{\alpha}}_{\nu\mu}) - \{{^{\varphi}}_{\alpha\mu}\}\,{{\cal
    K}^{\alpha}}_{\nu\varphi} - \{{^{\alpha}}_{\nu\varphi}\}\,{{\cal
    K}^{\varphi}}_{\alpha\mu}\Big),
\end{eqnarray}
where we have denoted 
\begin{eqnarray}\label{eq:6}
{\cal C} = g^{\mu\nu}\,{\cal C}_{\mu\nu} = g^{\mu\nu}({{\cal
    K}^{\varphi}}_{\alpha\mu}\,{{\cal K}^{\alpha}}_{\nu\varphi} -
{{\cal K}^{\alpha}}_{\alpha\varphi}{{\cal K}^{\varphi}}_{\nu\mu}).
\end{eqnarray}
In Eq.(\ref{eq:5}) removing the total derivatives and integrating by
parts we may delete the third term and transcribe the fourth term into
the form $\sqrt{-g}\,{g^{\mu\nu}}_{;\alpha}\,{{\cal
    K}^{\alpha}}_{\nu\mu}$, where ${g^{\mu\nu}}_{;\alpha}$ is the
covariant derivative of the metric tensor $g^{\mu\nu}$, vanishing
because of the metricity condition ${g^{\mu\nu}}_{;\alpha} = 0$. Thus,
the Einstein--Hilbert action Eq.(\ref{eq:1}) of the Einstein--Cartan
gravitational theory with the scalar curvature Eq.(\ref{eq:2}) can be
represented in the following additive form
\begin{eqnarray}\label{eq:7}
S_{\rm EH} = \frac{1}{2}\,M^2_{\rm Pl}\int d^4x\,\sqrt{-g}\,R +
\frac{1}{2}\,M^2_{\rm Pl}\int d^4x\,\sqrt{-g}\,{\cal C}.
\end{eqnarray}
Below we use the Einstein--Hilbert action Eq.(\ref{eq:7}) for the
derivation of the Einstein equations in the Einstein--Cartan
gravitational theory with the chameleon (quintessence) field, spinless
matter and torsion as an external (or a background) field
\cite{Shapiro2002,Kostelecky2004,Rumpf1979,Laemmerzahl1997,Kostelecky2008,Obukhov2014,Lehnert2014,Ivanov2015,Ivanov2015a,Ivanov2015b,Ivanov2016}.

\section{Einstein's equations in the Einstein--Cartan gravity with  
chameleon field and spinless matter}
\label{sec:torsion}

\subsection{Einstein's equations and torsion energy--momentum tensor}

Using Eq.(\ref{eq:7}) the action of the Einstein--Cartan gravity with
torsion and chameleon fields coupled to spinless matter we take in the
form
\begin{eqnarray}\label{eq:8}
S_{\rm EH} = \frac{1}{2}\,M^2_{\rm Pl}\,\int d^4x\,\sqrt{-g}\,R +
\frac{1}{2}\,M^2_{\rm Pl}\int d^4x\,\sqrt{-g}\,{\cal C} + \int
d^4x\,\sqrt{-g}\,{\cal L}[\phi] + \int d^4x\,\sqrt{- \tilde{g}}\,{\cal
  L}_m[\tilde{g}],
\end{eqnarray}
where ${\cal L}[\phi]$ is the Lagrangian of the chameleon field
\begin{eqnarray}\label{eq:9}
{\cal L}[\phi] =
\frac{1}{2}\,g^{\mu\nu}\,\partial_{\mu}\phi\partial_{\nu}\phi -
V(\phi),
\end{eqnarray}
where $V(\phi)$ is the potential of the chameleon self--interaction. A
spinless matter is described by the Lagrangian ${\cal
  L}_m[\tilde{g}_{\mu\nu}]$. The interaction of spinless matter with
the chameleon field runs through the metric tensor
$\tilde{g}_{\mu\nu}$ in the Jordan frame
\cite{Chameleon1,Chameleon2,Dicke1962}, which is conformally related
to the Einstein--frame metric tensor $g_{\mu\nu}$ by
$\tilde{g}_{\mu\nu} = f^2\,g_{\mu\nu}$ (or $\tilde{g}^{\mu\nu} =
f^{-2}\,g^{\mu\nu}$) and $\sqrt{- \tilde{g}} = f^4 \,\sqrt{-g}$ with
$f = e^{\,\beta\phi/M_{\rm Pl}}$, where $\beta$ is the
chameleon--matter coupling constant \cite{Chameleon1,Chameleon2}. The
factor $f = e^{\,\beta\phi/M_{\rm Pl}}$ can be interpreted also as a
conformal coupling to matter \cite{Dicke1962} (see also
\cite{Chameleon1,Chameleon2} and \cite{Ivanov2015}). Varying the
action Eq.(\ref{eq:8}) with respect to the metric tensor $\delta
g^{\mu\nu}$ (see, for example, \cite{Rebhan2012}) we arrive at the
Einstein equations, modified by the contribution of the chameleon
field and torsion
\begin{eqnarray}\label{eq:10}
R_{\mu\nu} - \frac{1}{2}\,g_{\mu\nu}\,R = - \frac{1}{M^2_{\rm Pl}}T_{\mu\nu},
\end{eqnarray}
where the Ricci tensor $R_{\mu\nu}$ and the scalar curvature $R$ are
expressed in terms of the Christoffel symbols only
$\{{^{\alpha}}_{\mu\nu}\}$ and the metric tensor $g_{\mu\nu}$ in
the Einstein frame \cite{Rebhan2012}. Then, $T_{\mu\nu}$ is the tensor
\begin{eqnarray}\label{eq:11}
T_{\mu\nu} = T^{(\phi)}_{\mu\nu} + f\,T^{(m)}_{\mu\nu} + T^{(\rm
  tors)}_{\mu\nu},
\end{eqnarray}
which can be identified with the energy--momentum tensor of the
torsion--chameleon--matter system, where $T^{(\phi)}_{\mu\nu}$ and
$T^{(m)}_{\mu\nu}$ are the chameleon field and matter (dark and baryon
matter) energy--momentum tensors. As has been shown in
\cite{Ivanov2016a}, the matter energy--momentum tensor
$T^{(m)}_{\mu\nu}$ appears in the right--hand--side (r.h.s.) of the
Einstein equations multiplied by the conformal factor $f$. In the CDM
model, accepted for the description of spinless matter in our analysis
of the Einstein--Cartan gravitational theory, the energy--momentum
tensor $T^{(m)}_{\mu\nu}$ has only time--time component $T^{(m)}_{00}
= \rho$, where $\rho$ is a spinless matter density in the Einstein
frame. In turn, the energy--momentum tensor $T^{(\phi)}_{\mu\nu}$ of
the scalar field is defined by
\begin{eqnarray}\label{eq:12}
T^{(\phi)}_{\mu\nu} = \frac{2}{\sqrt{- g}}\,\frac{\delta }{\delta
  g^{\mu\nu}}\Big(\sqrt{- g}\,{\cal L}[\phi]\Big) = \partial_{\mu}\phi
\partial_{\nu}\phi -
g_{\mu\nu}\,\Big(\frac{1}{2}\,g^{\lambda\rho}\,\partial_{\lambda}\phi\,
\partial_{\rho}\phi - V(\phi)\Big).
\end{eqnarray}
Then, the tensor $T^{(\rm tors)}_{\mu\nu}$ is caused by the
contribution of the torsion field and defined by
\begin{eqnarray}\label{eq:13}
T^{(\rm tors)}_{\mu\nu} &=& \frac{M^2_{\rm Pl}}{\sqrt{-
    g}}\,\frac{\delta }{\delta g^{\mu\nu}}\Big(\sqrt{-g}\,{\cal
  C}\Big).
\end{eqnarray}
We identify this tensor with the torsion energy--momentum tensor. The
properties of this tensor we investigate below. Now we would like to
rewrite the energy--momentum tensor of the scalar field in terms of
the energy momentum tensor of the chameleon one. For this aim we have
to take into account the equation of motion for the chameleon field
\cite{Ivanov2015}
\begin{eqnarray}\label{eq:14}
\frac{1}{\sqrt{-g}}\,\partial_{\mu}\Big(\sqrt{-
  g}\,\partial^{\mu}\phi\Big) + \frac{\partial V_{\rm
    eff}(\phi)}{\partial \phi} = 0,
\end{eqnarray}
where $V_{\rm eff}(\phi)$ is the effective potential for the chameleon
field given by \cite{Chameleon1,Chameleon2,Ivanov2016a}
\begin{eqnarray}\label{eq:15}
V_{\rm eff}(\phi) = V(\phi) + \rho\,(f(\phi) - 1),
\end{eqnarray}
and to replace in Eq.(\ref{eq:12}) the potential $V(\phi)$ of
self--interaction of the scalar (chameleon) field by the effective
potential $V(\phi) = V_{\rm eff}(\phi) - \rho\,(f(\phi) - 1)$. As a
result, the first two terms in the total energy--momentum tensor
Eq.(\ref{eq:11}) become represented in the following form
\begin{eqnarray}\label{eq:16}
T^{(\phi)}_{\mu\nu} + f\,T^{(m)}_{\mu\nu} = T^{(\rm ch)}_{\mu\nu} +
\Theta^{(m)}_{\mu\nu},
\end{eqnarray}
where $T^{(\rm ch)}_{\mu\nu}$ is the energy--momentum tensor of the
chameleon field. It is defined by Eq.(\ref{eq:12}) with the
replacement $V(\phi) \to V_{\rm eff}(\phi)$. Then,
$\Theta^{(m)}_{\mu\nu}$ is the modified matter energy--momentum
tensor, given by
\begin{eqnarray}\label{eq:17}
\Theta^{(m)}_{\mu\nu} = f\,T^{(m)}_{\mu\nu} - g_{\mu\nu}\,\rho\,(f -
1).
\end{eqnarray}
Now we may proceed to the analysis of local properties of the Einstein
equations, i.e. the Einstein tensor $G_{\mu\nu} = R_{\mu\nu} -
\frac{1}{2}\,g_{\mu\nu}R$, and the total energy--momentum tensor
$T_{\mu\nu} = T^{(\rm ch)}_{\mu\nu} + \Theta^{(m)}_{\mu\nu} + T^{(\rm
  tors)}_{\mu\nu}$, respectively.

\subsection{Bianchi identity and local conservation of total 
energy--momentum tensor}

The important property of the left--hand--side (l.h.s.) of the
Einstein equations is that the Einstein tensor $G^{\mu\nu} =
R^{\mu\nu} - \frac{1}{2}\,g^{\mu\nu}\,R$ obeys the Bianchi identity
${G^{\mu\nu}}_{;\mu} = 0$ in a curved spacetime with an arbitrary
metric $g_{\mu\nu}$ \cite{Rebhan2012}. This implies that the r.h.s. of
the Einstein equations, i.e. the total energy--momentum tensor
$T^{\mu\nu}$, should also possess a vanishing covariant divergence,
i.e. ${T^{\mu\nu}}_{;\mu} = 0$. As we have shown in Appendix A, the
energy--momentum tensor of the chameleon field $T^{(\rm ch){\mu\nu}}$
possesses a vanishing covariant divergence ${T^{(\rm
    ch)\mu\nu}}_{;\mu} = 0$ in a curved spacetime with an arbitrary
metric $g_{\mu\nu}$. Since torsion is independent of the chameleon
field and matter, the torsion energy--momentum tensor $T^{(\rm
  tors)\mu\nu}$ and the matter energy--momentum tensor
$\Theta^{(m)\mu\nu}$ should fulfil the constraints 
\begin{eqnarray}\label{eq:18}
{T^{(\rm tors)\mu\nu}}_{;\mu} &=& \frac{1}{\sqrt{-
    g}}\,\partial_{\mu}\Big(\sqrt{-g}\,T^{(\rm tors)\mu\nu}\Big) +
\{{^{\nu}}_{\mu\lambda}\}\, T^{(\rm tors)\mu\lambda} =
0,\nonumber\\ {\Theta^{(m)\mu\nu}}_{;\mu} &=& \frac{1}{\sqrt{-
    g}}\,\partial_{\mu}\Big(\sqrt{-g}\,\Theta^{(m)\mu\nu}\Big) +
\{{^{\nu}}_{\mu\lambda}\}\, \Theta^{(m)\mu\lambda} = 0
\end{eqnarray}
independently of each other.  As has been shown in \cite{Ivanov2016a},
local conservation of the matter energy--momentum tensor leads to the
evolution equation for the matter density.  Since in the CDM model,
which we accept here for the description of matter, the matter
energy--momentum tensor $\Theta^{(m)\mu\nu}$ is equal to
\begin{eqnarray}\label{eq:19}
\Theta^{(m)\mu\nu} = f\,\rho\,g^{\mu 0}g^{\nu 0} - \rho\,(f -
1)\,g^{\mu\nu},
\end{eqnarray}
the evolution equation for the matter density $\rho$ in a curved
spacetime with an arbitrary metric $g_{\mu\nu}$ is
\begin{eqnarray}\label{eq:20}
\frac{1}{\sqrt{-g}}\,\partial_{\mu}\Big(\sqrt{-g}\,f \rho\,g^{\mu
  0}g^{\nu 0} \Big) + \{{^{\nu}}_{\mu\lambda}\}\,f \rho\,g^{\mu
  0}g^{\lambda 0} = g^{\mu\nu}\,\partial_{\mu}\Big(\rho\,(f - 1)\Big),
\end{eqnarray}
where we have used the metricity condition ${g^{\mu\nu}}_{;\mu} =
0$. Then, Eq.(\ref{eq:20}) can be rewritten in the more convenient
form
\begin{eqnarray}\label{eq:21}
\partial^{\nu}\rho + \Big(g^{\nu 0}\partial^0(f\rho) -
\partial^{\nu}(f\rho)\Big) + \Big(\frac{1}{\sqrt{-
    g}}\,\partial_{\mu}\Big(\sqrt{-g}\,g^{\mu 0}g^{\nu 0}\Big) +
\{{^{\nu}}_{\mu\lambda}\} g^{\mu 0} g^{\lambda 0}\Big)(f\rho) = 0.
\end{eqnarray}
In the Friedmann flat spacetime the evolution equation
Eq.(\ref{eq:21}) reduces to the form \cite{Ivanov2016a}
\begin{eqnarray}\label{eq:22}
\dot{\rho} + 3\,{\rm H}\,\rho f = 0,
\end{eqnarray}
where ${\rm H} = \dot{a}/a$ is the Hubble rate. Now we may proceed to
the analysis of local conservation of the torsion energy--momentum
tensor $T^{(\rm tors)\mu\nu}$.

\subsection{Local conservation of torsion energy--momentum tensor}

Since torsion is an external  field, which does not obey any equation
of motion and boundary conditions, the requirement of local
conservation of the torsion energy--momentum tensor in a curved
spacetime with an arbitrary metric tensor can be fulfilled if and only
if the torsion energy--momentum tensor is proportional to a metric
tensor $T^{(\rm tors)\mu\nu} \sim g^{\mu\nu}$. In this case local
conservation of the torsion energy--momentum tensor ${T^{(\rm
    tors)\mu\nu}}_{;\mu} = 0$ is caused by the metricity condition
${g^{\mu\nu}}_{;\lambda} = 0$ \cite{Rebhan2012}, which is valid in the
Einstein--Cartan gravitational theory under consideration
\cite{Hehl1976}. Thus, we may set the torsion energy--meomentum tensor
equal to
\begin{eqnarray}\label{eq:23}
T^{(\rm tors)}_{\mu\nu} = \Lambda_C M^2_{\rm Pl}\,g_{\mu\nu} = - p_{\rm tors}\,g_{\mu\nu},
\end{eqnarray}
where $\Lambda_C$ is cosmological constant and $p_{\rm tors} = -
\Lambda_C M^2_{\rm Pl}$ can be interpreted as a torsion
pressure. According to the standard definition of the ``matter''
energy--momentum tensor \cite{Rebhan2012}, if the torsion
energy--momentum tensor is defined by Eq.(\ref{eq:23}) torsion obeys
the equation of state $\rho_{\rm tors} = - p_{\rm tors}$, where
$\rho_{\rm tors}$ is a torsion denisty in agreement with the
properties of dark energy \cite{Peebles2003,Copeland2006}. This gives
the following equation for ${\cal C}$
\begin{eqnarray}\label{eq:24}
\frac{M^2_{\rm Pl}}{\sqrt{- g}}\,\frac{\delta }{\delta
  g^{\mu\nu}}\Big(\sqrt{-g}\,{\cal C}\Big) = \Lambda_C M^2_{\rm Pl}\,g_{\mu\nu}.
\end{eqnarray}
Solving this equation we obtain
\begin{eqnarray}\label{eq:25}
{\cal C} = g^{\mu\nu}\,{\cal C}_{\mu\nu} = g^{\mu\nu}({{\cal
    K}^{\varphi}}_{\alpha\mu}\,{{\cal K}^{\alpha}}_{\nu\varphi} -
{{\cal K}^{\alpha}}_{\alpha\varphi}{{\cal K}^{\varphi}}_{\nu\mu}) = -
2\,\Lambda_C,
\end{eqnarray}
where we have used Eq.(\ref{eq:6}).  Cosmological constant $\Lambda_C$
is related to the relative dark energy density at our time as follows
$\Lambda_C = 3{\rm H}^2_0\Omega_{\Lambda}$, where ${\rm H}_0 =
1.437(26) \times 10^{-33}\,{\rm eV}$ and $\Omega_{\Lambda} \simeq
0.685$ are the Hubble constant and the relative dark energy density at
our time \cite{PDG2014}.

The relation Eq.(\ref{eq:25}) can be treated as a surface in the
24--dimensional space of torsion tensor field ${\cal
  T}_{\sigma\mu\nu}$ components, where the raising and lowering of
indices are performed with the metric tensors $g^{\mu\nu}$ and
$g_{\mu\nu}$, respectively.

\section{Conclusive discussion}
\label{sec:conclusion}

We have analysed the Einstein--Cartan gravitational theory in the
standard model--independent form ${\cal R} = R + {\cal K}^2$, where
$R$ and ${\cal K}^2$ are the contributions of the Einstein gravity and
torsion, respectively. We have extended also the Einstein--Cartan
gravity by the contribution of the chameleon (quintessence) field and
spinless matter (dark and baryon matter), described in the CDM model
in terms of a matter density $\rho$ in the Einstein frame. We have
added the chameleon field and spinless matter because of their
important role in the evolution of the Universe
\cite{Brax2004,Ivanov2016a}. We have shown that i) torsion does not
couple to a spinless matter and ii) the contribution of torsion to the
Einstein equations one may interpret in terms of the torsion
energy--momentum tensor as a part of the total energy--momentum tensor
$T^{\mu\nu} = T^{(\rm ch)\mu\nu} + \Theta^{(m)\mu\nu} + T^{(\rm
  tors)\mu\nu}$ of the system, including the chameleon field $T^{(\rm
  ch)\mu\nu}$, spinless matter $\Theta^{(m)\mu\nu}$ and torsion
$T^{(\rm tors)\mu\nu}$. The important property of the total
energy--momentum tensor is its local conservation, which is equivalent
to a vanishing covariant divergence ${T^{\mu\nu}}_{;\mu} = 0$ as a
consequence of the Bianchi identity ${G^{\mu\nu}}_{;\mu} = 0$ for the
Einstein tensor $G^{\mu\nu} = R^{\mu\nu} -
\frac{1}{2}\,g^{\mu\nu}\,R$. Since the Bianchi identity
${G^{\mu\nu}}_{;\mu} = 0$ is valid in a curved spacetime with an
arbitrary metric tensor $g_{\mu\nu}$ or an arbitrary gravitational
field \cite{Rebhan2012}, the total energy--momentum tensor
$T^{\mu\nu}$ should fulfil the constraint ${T^{\mu\nu}}_{;\mu} = 0$
also in a curved spacetime with an arbitrary metric tensor. We have
shown (see Appendix A) that the energy--momentum tensor of the
chameleon field fulfils the constraint ${T^{(\rm ch)\mu\nu}}_{;\mu} =
0$ identically for arbitrary metric. Then, the constraint
${\Theta^{(m)\mu\nu}}_{;\mu} = 0$ is equivalent to the evolution
equation of a matter. In the CDM model and in the Friedmann flat
spacetime such an evolution equation reduces to the evolution equation
of a pressureless matter density $\dot{\rho} + 3\,{\rm H}\,\rho\,f =
0$, which has been recently derived and analysed in
\cite{Ivanov2016a}, where ${\rm H}$ is the Hubble rate. As has been
discussed in \cite{Ivanov2016a} the presence of the conformal factor
$f$ in the evolution equation testifies an important role of the
chameleon field in a matter evolution in the Universe, during its
expansion. The traces of such an influence may be found in the
Cosmological Microwave Background (CMB) \cite{Ivanov2016a}. The local
properties of the energy--momentum tensors of the chameleon field and
spinless matter imply that the torsion energy--momentum tensor
$T^{(\rm tors)\mu\nu}$ should also possess a vanishing covariant
divergence ${T^{(\rm tors)\mu\nu}}_{;\mu} = 0$. Moreover, such a
covariant divergence should vanish in a curved spacetime with an
arbitrary metric tensor. Since torsion does not obey any equation of
motion and boundary conditions, the only one possibility to fulfil the
constraint ${T^{(\rm tors)\mu\nu}}_{;\mu} = 0$ is to set $T^{(\rm
  tors)\mu\nu} \sim g^{\mu\nu}$. In this case the constraint ${T^{(\rm
    tors)\mu\nu}}_{;\mu} = 0$ is fulfilled identically because of the
metricity condition ${g^{\mu\nu}}_{;\lambda} = 0$
\cite{Hehl1976,Rebhan2012}. Setting $T^{(\rm tors)}_{\mu\nu} =
\Lambda_C M^2_{\rm Pl}\,g_{\mu\nu}$, leading to the relation
Eq.(\ref{eq:25}) one may argue that torsion, serving as origin of
cosmological constant $\Lambda_C$, may explain a small value of
cosmological constant, which is a long--standing problem of cosmology
\cite{Weinberg1989,Peebles2003}. The relation Eq.(\ref{eq:25}) can be
interpreted as a surface in the 24--dimensional space of torsion
components.  It is obvious that the constraint Eq.(\ref{eq:25}) is not
very stringent and allows variations of torsion components in
sufficiently broad limits. Of course, any measurement of torsion
components is possible only through their interactions with spin
particles, for example, Dirac fermions
\cite{Laemmerzahl1997,Kostelecky2008,Obukhov2014,Ivanov2015a,Ivanov2015b,Ivanov2016}. As
has been shown in \cite{Ivanov2016}, in the curved spacetimes with
rotation one may, in principle, to observe all torsion components
through low--energy torsion--fermion effective potentials. However,
some low--energy torsion--fermion interactions are not defined by a
torsion--spin--fermion couplings (see Eq.(22) of
Ref.\cite{Ivanov2016}). As has been shown by Lehnert {\it et al.}
\cite{Lehnert2014}, cold neutrons can be a good tool for measurements
of torsion--spin--fermion interactions. As has been also discussed in
\cite{Ivanov2015b,Ivanov2016}, the qBounce experiments can provide a
precision analysis of all torsion--neutron low--energy interactions at
the level of sensitivities $\Delta E \sim (10^{-17} - 10^{-21})\,{\rm
  eV}$ \cite{Abele2010}.

According to Kostelecky \cite{Kostelecky2004}, torsion, treated as an
external (or a background) field, should be responsible for violation
of local Lorentz invariance or CPT invariance
\cite{Kostelecky1997,Kostelecky1998,Kostelecky2009}. A proportionality
of the torsion energy--momentum tensor to a metric tensor, required by
local conservation in a curved spacetime with an arbitrary metric
tenor, should be of use to avoid a no--go issue with the Bianchi
identities discovered in \cite{Kostelecky2004}. In effect, fixing
torsion to a background value may mean that torsion tensor components
should behave like Standard--Model Extension (SME) coefficients for
Lorentz violation, so their couplings to any matter or forces are
constrained by the various searches for Lorentz violation reported in
\cite{Kostelecky2008a}.

An attempt to relate cosmological constant to torsion has been
undertaken by Pop\l{}awski \cite{Poplawski2011,Poplawski2013}. In the
Einstein--Cartan gravitational theory with the Dirac--quark fields
Pop\l{}awski has varied the Einstein--Hilbert action with respect to
the contorsion tensor and replaced the torsion--Dirac--quark
interactions by the four--quark axial--vector--axial--vector
interaction, which he has equated with cosmological
constant. According to Pop\l{}awski \cite{Poplawski2011}, the vacuum
expectation value of such a four--quark interaction should correspond
cosmological constant, whereas spacetime fluctuations of the quark
fields should describe its spacetime dependence.  However, as has been
pointed out by Pop\l{}awski \cite{Poplawski2013}, the value of
cosmological constant, defined by the quark condensate
\cite{Poplawski2011}, is by a factor 8 larger compared to the
observable one \cite{PDG2014}.  Thus, in comparison with our result
the analysis of torsion--induced cosmological constant, proposed by
Pop\l{}awski \cite{Poplawski2011}, seems to be a model--dependent,
which does not reproduce the observable value of cosmological
constant. The references to other dynamical approaches for the
description of cosmological constant one may find in the papers by
Pop\l{}awski \cite{Poplawski2011,Poplawski2013}. The discussion of
these approaches goes beyond the scope of our paper.

Finally we would like to discuss the results, given in Appendix B,
where we have analysed the Poincar\'{e} gauge gravitational theory
\cite{Kibble1961} (see also
\cite{Utiyama1956,Sciama1961,Blagojevic2001} and
\cite{Hehl1976,Hehl2007,Hehl2012,Hehl2013,Obukhov2014}). We have shown
that the integrand of the Einstein--Hilbert action $e \,{\cal R} =
e\,{e^{\mu}}_a {e^{\nu}}_b {{\cal R}_{\mu\nu}}^{ab}$ of the
Poincar\'{e} gauge gravitational theory, where $e = \sqrt{- g}$ and
${{\cal R}_{\mu\nu}}^{ab}$ is the gravitational field strength tensor
of the Poincar\'{e} gauge gravitational theory, defined in terms of
the vierbein fields ${e^{\mu}}_a$ and ${e^{\nu}}_b$ and torsion, can
be represented in the additive form $e\,(R + {\cal C})$, where $R =
{e^{\mu}}_a {e^{\nu}}_b {R_{\mu\nu}}^{ab}$ and ${R_{\mu\nu}}^{ab}$ is
the gravitational field strength tensor of the Poincar\'{e} gauge
gravitational theory, defined only in terms of vierbein fields, and
${\cal C} = {{\cal K}^{\varphi}}_{\alpha\mu}\,{{\cal
    K}^{\alpha\mu}}_{\varphi} - {{\cal
    K}^{\alpha}}_{\alpha\varphi}{{\cal K}^{\varphi\mu}}_{\mu}$. This
allows to get a contribution of torsion to the Einstein equations in
the form of the torsion energy--momentum. A requirement of local
conservation of the torsion energy--momentum imposes its
proportionality to a metric tensor in complete agreement with the
result, obtained in the Einstein--Cartan gravitational theory
discussed in this paper.

 \section{Acknowledgements}

We thank Hartmut Abele for interest to our work. We are grateful to
Friedrich Hehl for interesting discussions and critical comments and
to Alan Kostelecky for fruitful encouraging discussions. This work was
supported by the Austrian ``Fonds zur F\"orderung der
Wissenschaftlichen Forschung'' (FWF) under the contracts I689-N16,
I862-N20 and P26781-N20.

\section{Appendix A: Analysis of local conservation of the 
energy--momentum tensor of the scalar field}
\renewcommand{\theequation}{A-\arabic{equation}}
\setcounter{equation}{0}

In this Appendix we calculate the covariant divergence of the
energy--momentum tensor of the chameleon field $T^{(\rm
  ch)^{\mu\nu}}$, defined by
\begin{eqnarray}\label{eq:A.1}
T^{(\rm ch)\mu\nu} = \frac{\partial \phi}{\partial
  x_{\mu}}\frac{\partial \phi}{\partial x_{\nu}} - g^{\mu\nu}\,{\cal
  L}^{(\rm ch)}_{\rm eff}[\phi],
\end{eqnarray}
where we have denoted 
\begin{eqnarray}\label{eq:A.2}
{\cal L}^{(\rm ch)}_{\rm eff}[\phi] =
\frac{1}{2}\,g^{\alpha\beta}\,\frac{\partial \phi}{\partial
  x^{\alpha}}\frac{\partial \phi}{\partial x^{\beta}} - V_{\rm
  eff}(\phi).
\end{eqnarray}
The requirement of local conservation of the energy--momentum tensor of
the chameleon field demands a vanishing covariant divergence
\begin{eqnarray}\label{eq:A.3}
{T^{(\rm ch)\mu\nu}}_{;\mu} = \frac{1}{\sqrt{-
    g}}\,\frac{\partial}{\partial x^{\rho}}\Big(\sqrt{-g}\,T^{(\rm
  ch)\rho\nu}\Big) + \{{^{\nu}}_{\mu\rho}\}\, T^{(\rm ch)\mu\rho} = 0.
\end{eqnarray}
Using the equation of motion for the chameleon field
\begin{eqnarray}\label{eq:A.4}
\frac{1}{\sqrt{-g}}\,\frac{\partial}{\partial
  x^{\mu}}\Big(\sqrt{-g}\,g^{\mu\nu}\frac{\partial \phi}{\partial
  x^{\nu}}\Big) + \frac{\partial V_{\rm eff}(\phi)}{\partial \phi} = 0
\end{eqnarray}
the calculation of the covariant divergence of the energy--momentum
tensor of the chameleon field runs as follows 
\begin{eqnarray}\label{eq:A.5}
\hspace{-0.3in}{T^{(\rm ch)\mu\nu}}_{;\mu} &=& \frac{1}{\sqrt{-
    g}}\,\frac{\partial}{\partial
  x^{\rho}}\Big(\sqrt{-g}\Big(\frac{\partial \phi}{\partial
  x_{\rho}}\,\frac{\partial \phi}{\partial x_{\nu}} -
g^{\rho\nu}\,{\cal L}^{(\rm ch)}_{\rm eff}[\phi]\Big) +
\{{^{\nu}}_{\mu\rho}\}\,\Big( \frac{\partial \phi}{\partial
  x_{\mu}}\,\frac{\partial \phi}{\partial x_{\rho}} -
g^{\mu\rho}\,{\cal L}^{(\rm ch)}_{\rm eff}[\phi]\Big)
=\nonumber\\ \hspace{-0.3in}&=&\frac{1}{\sqrt{-
    g}}\,\frac{\partial}{\partial
  x^{\rho}}\Big(\sqrt{-g}\Big(\frac{\partial \phi}{\partial
  x_{\rho}}\Big)\,\frac{\partial \phi}{\partial x_{\nu}} +
\frac{\partial \phi}{\partial x_{\rho}}\frac{\partial^2 \phi}{\partial
  x^{\rho}\partial x_{\nu}} - \frac{1}{\sqrt{-
    g}}\,\frac{\partial}{\partial
  x^{\rho}}\Big(\sqrt{-g}g^{\rho\nu}\Big)\,{\cal L}^{(\rm ch)}_{\rm
  eff}[\phi] - g^{\rho\nu}\frac{\partial {\cal L}^{(\rm ch)}_{\rm
    eff}[\phi]}{\partial
  x^{\rho}}\nonumber\\ \hspace{-0.3in}&-&\{{^{\nu}}_{\mu\rho}\}\,
g^{\mu\rho}\,{\cal L}[\phi] + \{{^{\nu}}_{\mu\rho}\}\,\frac{\partial
  \phi}{\partial x_{\mu}}\,\frac{\partial \phi}{\partial x_{\rho}} =
\frac{1}{\sqrt{- g}}\,\frac{\partial}{\partial
  x^{\rho}}\Big(\sqrt{-g}\Big(\frac{\partial \phi}{\partial
  x_{\rho}}\Big)\,\frac{\partial \phi}{\partial x_{\nu}} +
\frac{\partial \phi}{\partial x_{\rho}}\frac{\partial^2 \phi}{\partial
  x^{\rho}\partial x_{\nu}}\nonumber\\ &-& \frac{1}{\sqrt{-
    g}}\,\frac{\partial}{\partial
  x^{\rho}}\Big(\sqrt{-g}g^{\rho\nu}\Big)\,{\cal L}^{(\rm ch)}_{\rm
  eff}[\phi] - \frac{\partial}{\partial
  x_{\nu}}\Big(\frac{1}{2}\,g_{\mu\rho}\,\frac{\partial \phi}{\partial
  x_{\mu}}\frac{\partial \phi}{\partial x_{\rho}}\Big) +
\frac{\partial V_{\rm eff}(\phi)}{\partial \phi}\frac{\partial
  \phi}{\partial x_{\nu}} + \frac{1}{\sqrt{-
    g}}\,\frac{\partial}{\partial
  x^{\rho}}\Big(\sqrt{-g}g^{\rho\nu}\Big)\,{\cal L}^{(\rm ch)}_{\rm
  eff}[\phi]\nonumber\\ \hspace{-0.3in}&+&
\{{^{\nu}}_{\mu\rho}\}\,\frac{\partial \phi}{\partial
  x_{\mu}}\,\frac{\partial \phi}{\partial x_{\rho}},
\end{eqnarray}
where we have used the relation \cite{Rebhan2012}
\begin{eqnarray}\label{eq:A.6}
g^{\mu\nu}\,\{{^{\varphi}}_{\mu\nu}\} = - \frac{1}{\sqrt{-
    g}}\,\frac{\partial}{\partial x^{\lambda}}\Big(\sqrt{-
  g}\,g^{\varphi\lambda}\Big).
\end{eqnarray}
Cancelling like terms and using
Eq.(\ref{eq:A.4}) we arrive at the expression
\begin{eqnarray}\label{eq:A.7}
{T^{(\rm ch)\mu\nu}}_{;\mu} &=&\frac{\partial \phi}{\partial
  x_{\rho}}\frac{\partial^2 \phi}{\partial x^{\rho}\partial x_{\nu}} -
\frac{\partial}{\partial x_{\nu}}\Big(\frac{1}{2}\,\frac{\partial
  \phi}{\partial x^{\rho}}\frac{\partial \phi}{\partial x_{\rho}}\Big)
+ \{{^{\nu}}_{\mu\rho}\}\,\frac{\partial \phi}{\partial
  x_{\mu}}\,\frac{\partial \phi}{\partial x_{\rho}}.
\end{eqnarray}
Because of the relation \cite{Rebhan2012}
\begin{eqnarray}\label{eq:A.8}
\{{^{\nu}}_{\mu\rho}\}\,\frac{\partial \phi}{\partial
  x_{\mu}}\,\frac{\partial \phi}{\partial x_{\rho}} =
\Big(\frac{\partial \phi}{\partial x_{\nu}}\Big)_{;\rho}\frac{\partial
  \phi}{\partial x_{\rho}} - \frac{\partial^2 \phi}{\partial x^{\rho}
  \partial x_{\nu}}\frac{\partial \phi}{\partial x_{\rho}}
\end{eqnarray}
we may transcribe the r.h.s. of Eq.(\ref{eq:A.7}) into the form
\begin{eqnarray}\label{eq:A.9}
{T^{(\rm ch)\mu\nu}}_{;\mu} &=& \Big(\frac{\partial \phi}{\partial
  x_{\nu}}\Big)_{;\rho}\frac{\partial \phi}{\partial x_{\rho}} -
\frac{\partial}{\partial x_{\nu}}\Big(\frac{1}{2}\,\frac{\partial
  \phi}{\partial x^{\rho}}\frac{\partial \phi}{\partial x_{\rho}}\Big)
= \Big(\frac{\partial \phi}{\partial
  x_{\nu}}\Big)_{;\rho}\frac{\partial \phi}{\partial x_{\rho}} -
\Big(\frac{\partial \phi}{\partial x^{\rho}}\Big)^{;\nu}\frac{\partial
  \phi}{\partial x_{\rho}} = \nonumber\\ &=&
g^{\nu\lambda}\Big\{\Big(\frac{\partial \phi}{\partial
  x^{\lambda}}\Big)_{;\rho} - \Big(\frac{\partial \phi}{\partial
  x^{\rho}}\Big)_{;\lambda}\Big\}\frac{\partial \phi}{\partial
  x_{\rho}},
\end{eqnarray}
where we have used the relation \cite{Rebhan2012}
\begin{eqnarray}\label{eq:A.10}
\frac{\partial}{\partial x_{\nu}}\Big(\frac{1}{2}\,\frac{\partial
  \phi}{\partial x^{\rho}}\frac{\partial \phi}{\partial x_{\rho}}\Big)
= \Big(\frac{1}{2}\,\frac{\partial \phi}{\partial
  x^{\rho}}\frac{\partial \phi}{\partial x_{\rho}}\Big)^{;\nu} =
\Big(\frac{\partial \phi}{\partial x^{\rho}}\Big)^{;\nu}\frac{\partial
  \phi}{\partial x_{\rho}}.
\end{eqnarray}
Since the covariant derivatives $(\partial_{\lambda}\phi)_{;\rho}$ and
$(\partial_{\rho}\phi)_{;\lambda}$ are equal, i.e.
$(\partial_{\lambda}\phi)_{;\rho} = (\partial_{\rho}\phi)_{;\lambda}$,
we get ${T^{(\rm ch)\mu\nu}}_{;\mu} = 0$. This confirms local
conservation of the energy--momentum tensor of the chameleon field in
a curved spacetime with an arbitrary metric tensor.

\section{Appendix B: Equivalence between the Einstein--Cartan 
gravitational theory, considered in this paper, and the 
Poincar\'{e} gauge gravitational theory}
\renewcommand{\theequation}{B-\arabic{equation}}
\setcounter{equation}{0}

In this Appendix we show that the Poincar\'{e} gauge gravitational
theory field strength tensor ${{\cal
    R}_{\mu\nu}}^{ab}$, expressed in terms of the
spin connection ${\omega_{\mu}}^{ab}$ (or local
Lorentz connection) \cite{Kibble1961} (see also \cite{Kostelecky2004})
\begin{eqnarray}\label{eq:B.1}
  {{\cal R}_{\mu\nu}}^{ab} = \partial_{\nu} {\omega_{\mu}}^{ab} -
  \partial_{\mu}{\omega_{\nu}}^{ab} + {{\omega_{\nu}}^{a}}_{c}\,
          {\omega_{\mu}}^{cb} - {{\omega_{\mu}}^{a}}_{c}\,
          {\omega_{\nu}}^{cb}
\end{eqnarray}
is related to the Riemannian curvature tensor ${{\cal
    R}^{\alpha}}_{\beta\mu\nu}$ of the Einstein--Cartan gravitational
theory as \cite{Kibble1961} (see also \cite{Kostelecky2004})
\begin{eqnarray}\label{eq:B.2}
{{\cal R}^{\alpha}}_{\beta\mu\nu} =
\partial_{\nu}{\Gamma^{\alpha}}_{\mu\beta} -
\partial_{\mu}{\Gamma^{\alpha}}_{\nu\beta} +
        {\Gamma^{\alpha}}_{\nu\varphi}{\Gamma^{\varphi}}_{\mu\beta} -
        {\Gamma^{\alpha}}_{\mu\varphi}{\Gamma^{\varphi}}_{\nu\beta}
\end{eqnarray} 
by the relation ${{\cal R}_{\mu\nu}}^{ab} =
{e_{\alpha}}^{a}\, e^{\beta b}\, {{\cal
    R}^{\alpha}}_{\beta\mu\nu}$, where ${e_{\alpha}}^{a}$
and $e^{\beta b}$ are the vierbein fields. The indices
$a = 0,1,2,3$ are in the Minkowski spacetime. The lowering
and raising of the indices $a$ one performs with the Minkowski metric
tensors $\eta_{ab}$ and
$\eta^{ab}$, respectively. In turn, the indices
$\mu = 0,1,2,3$ are in a curved spacetime and the lowering and raising
of the indices $\mu$ one performs with the metric tensors $g_{\mu\nu}$
and $g^{\mu\nu}$, respectively. For the derivation of the relation
${{\cal R}_{\mu\nu}}^{ab} =
{e_{\alpha}}^{a}\, e^{\beta b}\, {{\cal
    R}^{\alpha}}_{\beta\mu\nu}$ we define the spin affine connection as
\cite{Kostelecky2004} (see also \cite{Ivanov2015a})
\begin{eqnarray}\label{eq:B.3}
  {\omega_{\mu}}^{ab} &=& -
  \partial_{\mu}{e_{\lambda}}^{a}\,e^{\lambda\,b}
  +
  {\Gamma^{\alpha}}_{\mu\lambda}\,{e_{\alpha}}^{a}\,e^{\lambda\,b}
  \quad,\quad {\omega_{\nu}}^{ab} = -
  \partial_{\nu}{e_{\lambda}}^{a}\,e^{\lambda\,b}
  +
  {\Gamma^{\alpha}}_{\nu\lambda}\,{e_{\alpha}}^{a}\,e^{\lambda\,b},
  \nonumber\\ {{\omega_{\nu}}^{a}}_{c} &=& -
  \partial_{\nu}{e_{\rho}}^{a}\,{e^{\rho}}_{c} +
          {\Gamma^{\beta}}_{\nu\rho}\,{e_{\beta}}^{a}\,
          {e^{\rho}}_{c}\quad,\quad{\omega_{\mu}}^{cb}
          = -
          \partial_{\mu}{e_{\lambda}}^{c}\,e^{\lambda\,b}
          +
          {\Gamma^{\alpha}}_{\mu\kappa}\,{e_{\alpha}}^{c}\,
          e^{\kappa\,b}.
\end{eqnarray}
Plugging Eq.(\ref{eq:B.3}) into Eq.(\ref{eq:B.1}) we arrive at the
expression
\begin{eqnarray}\label{eq:B.4}
  {{\cal R}_{\mu\nu}}^{ab} &=&
  -\partial_{\nu}(\partial_{\mu}{e_{\lambda}}^{a}\,e^{\lambda\,b}) +
  \partial_{\nu}({e_{\alpha}}^{a}\,e^{\lambda\,b})\,
          {\Gamma^{\alpha}}_{\mu\lambda} +
          {e_{\alpha}}^{a}\,e^{\lambda\,b}\,\partial_{\nu}
          {\Gamma^{\alpha}}_{\mu\lambda},\nonumber\\ &&
          +\partial_{\mu}(\partial_{\nu}{e_{\lambda}}^{a}\,e^{\lambda\,b})
          - \partial_{\mu}({e_{\alpha}}^{a}\,e^{\lambda\,b})\,
          {\Gamma^{\alpha}}_{\nu\lambda} -
          {e_{\alpha}}^{a}\,e^{\lambda\,b}\,\partial_{\mu}
          {\Gamma^{\alpha}}_{\nu\lambda},\nonumber\\ && + [
            \partial_{\nu}{e_{\rho}}^{a}\,{e^{\rho}}_{c} -
                    {\Gamma^{\beta}}_{\nu\rho}\,{e_{\beta}}^{a}\,
                    {e^{\rho}}_{c}][
            \partial_{\mu}{e_{\lambda}}^{c}\,e^{\lambda\,b} -
                    {\Gamma^{\alpha}}_{\mu\kappa}\,{e_{\alpha}}^{c}\,
                    e^{\kappa\,b}]\nonumber\\ && - [
            \partial_{\mu}{e_{\rho}}^{a}\,{e^{\rho}}_{c} -
                    {\Gamma^{\beta}}_{\mu\rho}\,{e_{\beta}}^{a}\,
                    {e^{\rho}}_{c}][
            \partial_{\nu}{e_{\lambda}}^{c}\,e^{\lambda\,b} -
                    {\Gamma^{\alpha}}_{\nu\kappa}\,{e_{\alpha}}^{c}\,
                    e^{\kappa\,b}]
\end{eqnarray}
Using the properties of the vierbein fields \cite{Ivanov2015a} we get
${{\cal R}_{\mu\nu}}^{ab} =
{e_{\alpha}}^{a}\,e^{\beta b}\,{{\cal
    R}^{\alpha}}_{\beta\mu\nu} +
{O_{\mu\nu}}^{ab}$, where
${O_{\mu\nu}}^{ab}$ is defined by
\begin{eqnarray}\label{eq:B.5}
{O_{\mu\nu}}^{ab}&=& -
(\partial_{\nu}{e_{\lambda}}^{a})\,(\partial_{\mu}e^{\,\lambda b}) -
(\partial_{\mu}{e_{\alpha}}^{a})\,e^{\,\lambda
  b}\,{\Gamma^{\alpha}}_{\nu\lambda} -
{e_{\alpha}}^{b}\,(\partial_{\mu}e^{\,\lambda
  b})\,{\Gamma^{\alpha}}_{\nu\lambda} \nonumber\\ &&- (
\partial_{\mu}{e_{\lambda}}^{a})\,(\partial_{\nu}e^{\,\lambda b}) +
(\partial_{\nu}{e_{\alpha}}^{a})\,e^{\,\lambda
  b}\,{\Gamma^{\alpha}}_{\mu\lambda} +
        {e_{\alpha}}^{b}\,(\partial_{\nu}e^{\,\lambda
          b})\,{\Gamma^{\alpha}}_{\mu\lambda}
        \nonumber\\ &&-(\partial_{\mu}{e_{\rho}}^{a})\,{e^{\rho}}_{c}\,(\partial_{\nu}{e_{\lambda}}^{c})\,e^{\,\lambda
          b} +
                    {e_{\alpha}}^{a}\,{e^{\rho}}_{c}(\partial_{\nu}{e_{\lambda}}^{c})\,e^{\,\lambda
                      b}\,{\Gamma^{\alpha}}_{\mu\rho} +
                    (\partial_{\mu}{e_{\rho}}^{a})\,{e^{\rho}}_{c}\,{e_{\alpha}}^{c}\,e^{\,\kappa
                      b}\,{\Gamma^{\alpha}}_{\nu\kappa}\nonumber\\ &&+
                    (\partial_{\nu}{e_{\rho}}^{a})\,{e^{\rho}}_c\,(\partial_{\mu}{e_{\lambda}}^c)\,e^{\,\lambda
                      b} -
                    {e_{\alpha}}^{a}\,{e^{\rho}}_{c}(\partial_{\mu}{e_{\lambda}}^{c})\,e^{\,\lambda
                      b}\,{\Gamma^{\alpha}}_{\nu\rho} -
                    (\partial_{\nu}{e_{\rho}}^{a})\,{e^{\rho}}_{c}\,{e_{\alpha}}^{c}\,e^{\,\kappa
                      b}\,{\Gamma^{\alpha}}_{\mu\kappa}.
\end{eqnarray}
Using the relations
${e^{\rho}}_{c}(\partial_{\alpha}{e_{\lambda}}^{c})
= -
(\partial_{\alpha}{e^{\rho}}_{c})\,{e_{\lambda}}^{c}$
and ${e_{\lambda}}^{c}\,e^{\,\lambda b} =
\eta^{cb}$ one may show that
${O_{\mu\nu}}^{ab} \equiv 0$. This gives
\begin{eqnarray}\label{eq:B.6}
  {{\cal R}_{\mu\nu}}^{ab} =
  {e_{\alpha}}^{a}\,e^{\beta b}\,{{\cal
      R}^{\alpha}}_{\beta\mu\nu} \quad,\quad {{\cal
      R}^{\alpha}}_{\beta\mu\nu} =
  {e^{\alpha}}_{a}\,e_{\beta b}\,{{\cal
      R}_{\mu\nu}}^{ab}.
\end{eqnarray}
Thus, we have confirmed the relations between the Riemannian curvature
tensor ${{\cal R}^{\alpha}}_{\beta\mu\nu}$ and the Poincar\'{e} gauge
gravitational field strength tensor ${{\cal
    R}_{\mu\nu}}^{ab}$, proposed for the first
time by Kibble \cite{Kibble1961} (see also \cite{Kostelecky2004}). The
relation Eq.(\ref{eq:10}) testifies the equivalence between the
Einstein--Cartan gravitational theory with the Riemannian curvature
tensor Eq.(\ref{eq:B.2}), defined in terms of the affine connection
Eq.(\ref{eq:3}), and the Poincar\'{e} gauge gravitational theory
\cite{Kibble1961} (see also \cite{Hehl1976,Hehl2013,Obukhov2014}) with
the Poincar\'{e} gauge gravitational field strength tensor
Eq.(\ref{eq:B.1}), defined in terms of the spin (or local Lorentz)
connection ${\omega_{\mu}}^{ab}$ and the vierbein
field ${e^{\mu}}_{a}$ and
${e_{\mu}}^{a}$. Indeed, the Einstein--Hilbert action
Eq.(\ref{eq:1}) can be written as follows \cite{Kostelecky2004}
\begin{eqnarray}\label{eq:B.7}
 S_{\rm EH} = \frac{1}{2}\,M^2_{\rm Pl}\int d^4x\,\sqrt{- g}\,{\cal R}
 = \frac{1}{2}\,M^2_{\rm Pl}\int
 d^4x\,e\,{e^{\mu}}_{a}{e^{\nu}}_{b}\,{{\cal
     R}_{\mu\nu}}^{ab},
\end{eqnarray}
where the Poincar\'{e} gauge gravitational field strength tensor
${{\cal R}_{\mu\nu}}^{ab}$ is given by Eq.(\ref{eq:B.1}) as a
functional of the spin connection ${\omega_{\mu}}^{ab}$ and the
vierbein fields ${e_{\mu}}^{a}$ and ${e^{\mu}}_{a}$,
respectively. Then, $e$ is the determinant $ e = {\rm
  det}\{{e_{\mu}}^{a}\}$, i.e. $\sqrt{- g} = \sqrt{ - {\rm
    det}\{g_{\mu\nu}\}} = \sqrt{ - {\rm
    det}\{\eta_{ab}{e_{\mu}}^{a}{e_{\nu}}^{b}\}} = e$. Now we may show
that the Einstein--Hilbert action Eq.(\ref{eq:B.7}) can be represented
in the additive form analogous to Eq.(\ref{eq:7}). For
this aim we define the spin affine connection ${\omega_{\mu}}^{ab}$ as
follows
\begin{eqnarray}\label{eq:B.8}
{\omega_{\mu}}^{ab} = {E_{\mu}}^{ab} + {{\cal K}_{\mu}}^{ab},
\end{eqnarray}
where ${E_{\mu}}^{ab}$ and ${{\cal K}_{\mu}}^{ab}$ are given by
\cite{Kostelecky2004}
\begin{eqnarray}\label{eq:B.9}
{E_{\mu}}^{ab} &=& \frac{1}{2}\,e^{\nu a}(\partial_{\mu}{e_{\nu}}^b -
\partial_{\nu}{e_{\mu}}^b) - \frac{1}{2}\,e^{\nu
  b}(\partial_{\mu}{e_{\nu}}^a - \partial_{\nu}{e_{\mu}}^a) -
\frac{1}{2}\,e^{\alpha a}\,e^{\beta
  b}\,{e_{\mu}}^c\,(\partial_{\alpha}e_{\beta c} -
\partial_{\beta}e_{\alpha c}),\nonumber\\ {{\cal K}_{\mu}}^{ab} &=&
        {\cal K}_{\alpha \mu \beta}\,e^{\alpha a}\,e^{\beta b}.
\end{eqnarray}
Plugging Eq.(\ref{eq:B.9}) into Eq.(\ref{eq:B.7}) we arrive at the
Einstein--Hilbert action
\begin{eqnarray}\label{eq:B.10}
 S_{\rm EH} &=&\frac{1}{2}\,M^2_{\rm Pl}\int
 d^4x\,e\,{e^{\mu}}_{a}{e^{\nu}}_{b}\,{{\cal R}_{\mu\nu}}^{ab} =
 \frac{1}{2}\,M^2_{\rm Pl}\int d^4x\,e\,R + \frac{1}{2}\,M^2_{\rm
   Pl}\int d^4x\,e\,{\cal C} + \bar{S}_{\rm EH},
\end{eqnarray}
where $R = {e^{\mu}}_{a}{e^{\nu}}_{b}\,{R_{\mu\nu}}^{ab}$ is the
functional of ${E_{\mu}}^{ab}$. It is defined only in terms of the
vierbein fields and corresponds to the contribution of the scalar
curvature in the Einstein gravity, whereas ${\cal C}$ is given by
${\cal C} = {{\cal K}^{\varphi}}_{\alpha\mu}\,{{\cal
    K}^{\alpha\mu}}_{\varphi} - {{\cal
    K}^{\alpha}}_{\alpha\varphi}{{\cal K}^{\varphi\mu}}_{\mu}$ and
corresponds to the contribution of torsion (see
Eq.(\ref{eq:6})). Then, the term $\bar{S}_{\rm EH}$ is equal to
\begin{eqnarray}\label{eq:B.11}
 \bar{S}_{\rm EH} = \frac{1}{2}\,M^2_{\rm
   Pl}\int
 d^4x\,e\,{e^{\mu}}_{a}{e^{\nu}}_{b}\,\Big(\partial_{\nu}{{\cal
     K}_{\mu}}^{ab} - \partial_{\mu}{{\cal K}_{\nu}}^{ab} +
 {{E_{\nu}}^a}_c {{\cal K}_{\mu}}^{cb} + {E_{\mu}}^{cb}\,{{{\cal
       K}_{\nu}}^a}_c - {{E_{\mu}}^a}_c\,{{\cal K}_{\nu}}^{cb} -
 {E_{\nu}}^{cb}\,{{{\cal K}_{\mu}}^a}_c\Big).
\end{eqnarray}
Below we show that $\bar{S}_{\rm EH} = 0$. The first step to the
realization of this aim is to define the Christoffel symbols in terms
of the vierbein fields. We get
\begin{eqnarray}\label{eq:B.12}
\{{^{\alpha}}_{\mu\nu}\} =
\frac{1}{2}\,{e^{\alpha}}_a(\partial_{\mu}{e_{\nu}}^a +
\partial_{\nu}{e_{\mu}}^a) + \frac{1}{2}\,{e^{\alpha}}_a\,e^{\beta
  a}(e_{\mu b}\partial_{\nu}{e_{\beta}}^b + e_{\nu
  b}\partial_{\mu}{e_{\beta}}^b) -
\frac{1}{2}\,{e^{\alpha}}_a\,e^{\beta a}(e_{\mu
  b}\partial_{\beta}{e_{\nu}}^b + e_{\mu
  b}\partial_{\beta}{e_{\mu}}^b).
\end{eqnarray}
Then, using the definitions for ${E_{\mu}}^{ab}$ and
$\{{^{\alpha}}_{\mu\nu}\}$, given by Eq.(\ref{eq:B.9}) and
Eq.(\ref{eq:B.12}), respectively, one may show that the covariant
derivative of the vierbein field ${{e_{\nu}}^a}_{;\mu}$ and
${e^{\nu}}_{a;\mu} $, defined by \cite{Kostelecky2004,Sciama1961}
\begin{eqnarray}\label{eq:B.13}
{e_{\nu}}^a_{;\mu} &=& \partial_{\mu}{e_{\nu}}^a -
\{{^{\alpha}}_{\mu\nu}\}\,{e_{\alpha}}^a +
  {{E_{\mu}}^a}_b\,{e_{\nu}}^b,\nonumber\\
{e^{\nu}}_{a;\mu} &=& \partial_{\mu} {e^{\nu}}_a  +
\{{^{\nu}}_{\rho \mu}\}\,{e^{\rho}}_a   +
  {E_{\mu a}}^b\,{e^{\nu}}_b,
\end{eqnarray}
are equal to zero, i.e. ${e_{\nu}}^a_{;\mu} = 0$ and
${e^{\nu}}_{a;\mu} = 0$. Integrating by parts in Eq.(\ref{eq:B.11}) we
arrive at the expression
\begin{eqnarray}\label{eq:B.14}
 \bar{S}_{\rm EH} &=& \frac{1}{2}\,M^2_{\rm Pl}\int d^4x\,\Big( - {{\cal
     K}_{\mu}}^{ab}\partial_{\nu}(e\,{e^{\mu}}_{a}{e^{\nu}}_{b}) +
     {{\cal
         K}_{\nu}}^{ab}\partial_{\mu}(e\,{e^{\mu}}_{a}{e^{\nu}}_{b}) +
     e\,{e^{\mu}}_{a}{e^{\nu}}_{b}\,{{E_{\nu}}^a}_c {{\cal
         K}_{\mu}}^{cb} +
     e\,{e^{\mu}}_{a}{e^{\nu}}_{b}\,{E_{\mu}}^{cb}\,{{{\cal
           K}_{\nu}}^a}_c\nonumber\\ &-&
     e\,{e^{\mu}}_{a}{e^{\nu}}_{b}\,{{E_{\mu}}^a}_c\,{{\cal
         K}_{\nu}}^{cb} -
     e\,{e^{\mu}}_{a}{e^{\nu}}_{b}\,{E_{\nu}}^{cb}\,{{{\cal
           K}_{\mu}}^a}_c\Big).
\end{eqnarray}
Calculating the first order derivatives we get
\begin{eqnarray}\label{eq:B.15}
 \bar{S}_{\rm EH} &=& \frac{1}{2}\,M^2_{\rm Pl}\int d^4x\,\Big( -
     {{\cal
         K}_{\mu}}^{ab}\,{e^{\mu}}_{a}\,\partial_{\nu}(e\,{e^{\nu}}_{b})
     - {{\cal
         K}_{\mu}}^{ab}\,e\,{e^{\nu}}_{b}\,\partial_{\nu}{e^{\mu}}_a +
     {{\cal
         K}_{\nu}}^{ab}\,{e^{\nu}}_{b}\,\partial_{\mu}(e\,{e^{\mu}}_a)
     + {{\cal
         K}_{\nu}}^{ab}\,e\,{e^{\mu}}_a\,\partial_{\mu}{e^{\nu}}_b\nonumber\\ &+&
     e\,{e^{\mu}}_{a}{e^{\nu}}_{b}\,{{E_{\nu}}^a}_c {{\cal
         K}_{\mu}}^{cb} +
     e\,{e^{\mu}}_{a}{e^{\nu}}_{b}\,{E_{\mu}}^{cb}\,{{{\cal
           K}_{\nu}}^a}_c -
     e\,{e^{\mu}}_{a}{e^{\nu}}_{b}\,{{E_{\mu}}^a}_c\,{{\cal
         K}_{\nu}}^{cb} -
     e\,{e^{\mu}}_{a}{e^{\nu}}_{b}\,{E_{\nu}}^{cb}\,{{{\cal
           K}_{\mu}}^a}_c\Big),
\end{eqnarray}
where we may combine some terms into the covariant divergences of the vierbein fields
\begin{eqnarray}\label{eq:B.16}
 \bar{S}_{\rm EH} &=& \frac{1}{2}\,M^2_{\rm Pl}\int d^4x\,\Big( -
     {{\cal
         K}_{\mu}}^{ab}\,{e^{\mu}}_{a}\,e\,{e^{\nu}}_{b;\nu} +
     {{\cal
         K}_{\nu}}^{ab}\,{e^{\nu}}_{b}\,e\,{e^{\mu}}_{a;\mu}
     - {{\cal
         K}_{\mu}}^{ab}\,e\,{e^{\nu}}_{b}\,\partial_{\nu}{e^{\mu}}_a 
     + {{\cal
         K}_{\nu}}^{ab}\,e\,{e^{\mu}}_a\,\partial_{\mu}{e^{\nu}}_b\nonumber\\ &+&
     e\,{e^{\mu}}_{a}{e^{\nu}}_{b}\,{{E_{\nu}}^a}_c {{\cal
         K}_{\mu}}^{cb} +
     e\,{e^{\mu}}_{a}{e^{\nu}}_{b}\,{E_{\mu}}^{cb}\,{{{\cal
           K}_{\nu}}^a}_c\Big).
\end{eqnarray}
Since ${e^{\nu}}_{b;\nu} = {e^{\mu}}_{a;\mu} = 0$, we get
\begin{eqnarray}\label{eq:B.17}
 \bar{S}_{\rm EH} = \frac{1}{2}\,M^2_{\rm Pl}\int d^4x\,\Big( -
     {{\cal
         K}_{\mu}}^{ab}\,e\,{e^{\nu}}_{b}\,\partial_{\nu}{e^{\mu}}_a +
     {{\cal K}_{\nu}}^{ab}\,e\,{e^{\mu}}_a\,\partial_{\mu}{e^{\nu}}_b
     + e\,{e^{\mu}}_{a}{e^{\nu}}_{b}\,{{E_{\nu}}^a}_c {{\cal
         K}_{\mu}}^{cb} +
     e\,{e^{\mu}}_{a}{e^{\nu}}_{b}\,{E_{\mu}}^{cb}\,{{{\cal
           K}_{\nu}}^a}_c\Big).
\end{eqnarray}
The integrand of Eq.(\ref{eq:B.17}) we rewrite as follows
\begin{eqnarray}\label{eq:B.18}
 \bar{S}_{\rm EH} &=& \frac{1}{2}\,M^2_{\rm Pl}\int d^4x\,\Big( -
     {{\cal
         K}_{\mu}}^{ab}\,e\,{e^{\nu}}_{b}\,(\partial_{\nu}{e^{\mu}}_a
     + {E_{\nu a}}^c\,{e^{\mu}}_c) + {{\cal
         K}_{\nu}}^{ab}\,e\,{e^{\mu}}_a\,(\partial_{\mu}{e^{\nu}}_b +
     {E_{\mu b}}^c\,{e^{\nu}}_c)\Big) =\nonumber\\ &=&M^2_{\rm Pl}\int
     d^4x\, {{\cal
         K}_{\mu}}^{ab}\,e\,{e^{\nu}}_{b}\,\{{^{\mu}}_{\rho\,
       \nu}\}\,{e^{\rho}}_a = M^2_{\rm Pl}\int d^4x\, e\,{{{\cal K}^{\rho}}_{\mu}}^{\nu}\,\{{^{\mu}}_{\rho\,
       \nu}\} = 0.
\end{eqnarray}
Thus, we have shown that $\bar{S}_{\rm EH} \equiv 0$. This means that
the Einstein--Hilbert action Eq.(\ref{eq:B.7}) can be written in the
additive form
\begin{eqnarray}\label{eq:B.19}
 S_{\rm EH} = \frac{1}{2}\,M^2_{\rm Pl}\int
 d^4x\,e\,{e^{\mu}}_{a}{e^{\nu}}_{b}\,{{\cal R}_{\mu\nu}}^{ab} =
 \frac{1}{2}\,M^2_{\rm Pl}\int d^4x\,e\,R + \frac{1}{2}\,M^2_{\rm
   Pl}\int d^4x\,e\,{\cal C},
\end{eqnarray}
where $R = {e^{\mu}}_{a}{e^{\nu}}_{b}\,{R_{\mu\nu}}^{ab}$ is defined
only in terms of the vierbein fields and corresponds to the
contribution of the scalar curvature in the Einstein gravity, whereas
${\cal C}$ is given by ${\cal C} = {{\cal
    K}^{\varphi}}_{\alpha\mu}\,{{\cal K}^{\alpha\mu}}_{\varphi} -
{{\cal K}^{\alpha}}_{\alpha\varphi}{{\cal K}^{\varphi\mu}}_{\mu}$ and
corresponds to the contribution of torsion (see Eq.(\ref{eq:6})). For
the derivation of Eq.(\ref{eq:B.19}) we have used the definition of
the covariant derivatives of the vierbein fields Eq.(\ref{eq:B.13})
and the properties of the contorsion tensor ${{\cal K}_{\mu}}^{ab} = -
{{\cal K}_{\mu}}^{ba}$ and ${\cal K}_{\alpha\mu\beta} = - {\cal
  K}_{\beta \mu\alpha}$ \cite{Kostelecky2004}.

The obtained result Eq.(\ref{eq:B.19}) testifies a complete
equivalence between the Einstein--Cartan gravitational theory,
analysed in this paper, and the Poincar\'{e} gauge gravitational
theory by Kibble \cite{Kibble1961} (see also
\cite{Sciama1961,Utiyama1956,Blagojevic2001} and
\cite{Hehl1976,Hehl2007,Hehl2012,Hehl2013,Obukhov2014}). This also
confirms the identification of the torsion contribution to the
Einstein equations with the torsion energy--momentum tensor
Eq.(\ref{eq:23}), local conservation of which can be reached only
through the relation Eq.(\ref{eq:24}), allowing to set ${\cal C} = -
2\Lambda_C$ (see Eq.(\ref{eq:25})).

\end{document}